\documentclass[twocolumn,showpacs,preprintnumbers,amsmath,amssymb,revsymb,prb, superscriptaddress]{revtex4}
\usepackage[dvips]{graphicx}
\usepackage{mathptmx}

\begin{document}
\title{Ab-initio formulation of the 4-point conductance of interacting
    electronic systems}
\author{P. Bokes } \email{peter.bokes@stuba.sk}
\affiliation{Department of Physics, University of York, Heslington, York
         YO10 5DD, United Kingdom}
\affiliation{Department of Physics, Faculty of Electrical Engineering and
        Information Technology, Slovak University of Technology,
    Ilkovi\v{c}ova 3, 812 19 Bratislava, Slovak Republic}
\author{J. Jung}
\affiliation{Physics Division, National Center for Theoretical Sciences,  
    P.O. Box 2-131, Hsinchu, Taiwan }
\affiliation{Department of Physics, University of York, Heslington, York
         YO10 5DD, United Kingdom}
\author{R. W. Godby}
\affiliation{Department of Physics, University of York, Heslington, York
         YO10 5DD, United Kingdom}

\date{\today{}}

\begin{abstract}
We derive an expression for the 4-point conductance of a general quantum
junction in terms of the density response function. Our formulation allows
us to show that the 4-point conductance of an interacting electronic system
possessing
either a geometrical constriction and/or an opaque barrier
becomes identical to the macroscopically measurable 2-point conductance.
Within time-dependent density-functional theory the formulation leads
to a direct identification of the functional form of the exchange-correlation
kernel that is important for the conductance. We demonstrate the practical implementation
of our formula for a metal-vacuum-metal interface.
\end{abstract}


\maketitle

\section{Introduction}
\label{sec-1}
Impressive progress has been achieved within the non-equilibrium Green's
function (NEGF) formulation of quantum transport using the simple
ground-state density-functional exchange-correlation potential
in a self-consistent formulation~\cite{Taylor02, Basch05} (NEGF-DFT).
However, limitations of the latter approximation were recently
identified~\cite{Sai05,Jung07,Toher05,Burke05,Palacios05,Stefanucci04,DiVentra04b,Koentopp07}.
For instance, NEGF-DFT's omission of the derivative-discontinuity in
the exchange-correlation energy functional was found responsible for serious
errors in transport calculation through localized resonant
levels~\cite{Toher05,Burke05}. Improvements through an (spin-) unrestricted
NEGF-DFT formulation have been argued to describe properly some aspects
of the Coulomb blockade in quantum junctions~\cite{Palacios05}. At this
level of the theory the exchange-correlation potential of the equilibrium
system, $v_{xc}$, is responsible for the electron interaction effects.

In a further theoretical development, Na Sai {\it et al.}~\cite{Sai05}
identified a dynamical correction to the resistance of a quantum junction
stemming from the contribution of the exchange-correlation electric field
to the overall drop in the total potential.
They estimated the correction within time-dependent current-density
functional theory~\cite{Vignale97} (TDCDFT) and showed that it has
its origin in the non-local density-dependence of the functional. The
very applicability of time-dependent density-functional theory (TDDFT)
to the problem of quantum transport in the long-time limit has been discussed
in depth by Stefanucci and Almbladh~\cite{Stefanucci04} and by
Di Ventra and Todorov~\cite{DiVentra04b}.

Several authors have proposed alternative treatments that avoid
the complexities of the exchange-correlation kernels of TD(C)DFT, either
by treating the central region with the configuration integration
method~\cite{Delaney04,Fagas06} while approximating the non-equilibrium
distribution of the electrons, or by using  the usual NEGF-DFT approach
in combination with a model self-energy within the central
region~\cite{Ferretti05}. More systematic approach to the self-energy
can be obtained through well tested approximations like the $GW$
method~\cite{Darancet07,Thygesen07}. Due to the large computational demand
so far only very small systems with restricted size of the basis set could
be studied. However, the results are encouraging, e.g. the Kondo effect
phenomenology seems to be well described within the self-consistent
$GW$ method~\cite{Thygesen07}.

Nonetheless, a systematic approach for addressing the conductance
of a fully interacting system at the {\it ab-initio} level is not available.
This is partly due to the fact that the very formulation of the NEGF
formalism~\cite{Meir92} is based on the concept of noninteracting electrodes
and demands partitioning of the system and the electron-electron
interaction~\cite{Prange63}. Stefanucci and Almbladh~\cite{Stefanucci04}
showed that the partitioning can be avoided in principle but
practical inclusion of the many-body interactions into the formalism
seems to be extremely cumbersome. Fortunately, partitioning
is not necessary within the linear response formulation. 
Several
authors have addressed the conductance of interacting system of electrons
within the framework of the Kubo formalism arriving at a 2-point 
Landauer-like formula for the 
conductance~\cite{Malet05,Prodan07,Bokes04,Koentopp07} by making certain 
assumptions about the steady-state total electric field.
There is, however, a
problem with this derivation since it ignores the charge redistribution
in the conducting system when the steady state is forming. In fact,
ignoring these aspects one quickly arrives at various unphysical corrections
to the conductance~\cite{Bokes_condmat06}. The problem of charge
redistributions has been first pointed out by Thouless~\cite{Thouless81}.
Later Kamenev and Kohn~\cite{Kamenev01} cast it into a self-consistent
framework for many-terminal conductance.

In our work we further develop
the formalism that treats the charge redistribution correctly, find
its physical interpretation in terms of a time-dependent transient process
that leads to the establishment of a current-carrying steady state, and derive
a closed formula for the 4-point conductance that is well-defined formally as well as physically.

To describe quantum transport it is  necessary to consider the production of
a steady state with nonzero current in a generic quantum junction,
starting from the ground state (or more generally, from an equilibrium
ensemble at finite temperature).
Physically, we expect to achieve a steady state that has a uniform average current
$I$ flowing through the system, accompanied by a
total potential exhibiting a drop $V$ across it.
As was advocated many years ago by Landauer~\cite{Landauer57},
$V$ arises from the resistivity dipole, a local charge imbalance
of a piled-up charge in the front and depleted charge behind a quantum
junction.
Physically such a state is reached by attaching a macroscopic battery
to the circuit that pumps electrons from one electrode to the other until a
certain potential difference between these two is established. For modeling
purposes this mechanism cannot be used directly~\footnote{Recently, clear 
signatures of the resistivity dipoles were found in time-dependent
simulations when starting from a charged initial state\cite{Sai07}.}. 
Here we will use a construct based on an auxiliary homogeneous vector 
potential $\vec{A}^{aux}(t)=
-\int_0^t \vec{E}^{aux}(t') dt'$, pointing in the direction of the eventual
flow of current, giving a momentum transfer to all the electrons in
the infinite system within a finite interval of time.
(This auxiliary
vector potential is similar in spirit to the one used by
R. Gebauer and R. Car to model a nanojunction using periodic boundary
conditions~\cite{Gebauer04b}. There, in contrast to our work, the vector
potential grows
linearly in time for all $t>0$ and the work exerted on the system must
be dissipated via auxiliary phonons located in the electrodes.)
The infinite extent of the system is formally essential to our treatment.
It is necessary for a continuous spectrum and for giving a momentum
transfer at $t=0$
to infinitely many electrons present in the system so that the current
will flow for all times $t>0$. However, we would like to mention already
at this point that in practical calculations the necessity of the infinite
extension can be relaxed so that the formulation can be used for practical
{\it ab-initio} calculations.

\section{The non-local conductivity and conductance}
\label{sec-2}
The response of the current density to a general external electric field 
is given within the linear response theory by~\cite{Kubo59}
\begin{eqnarray}
        \vec{j}(\vec{r},t) = \int_{-\infty}^{t} dt' \int d^3 r'
               \vec{\vec{\sigma}}(\vec{r},\vec{r}';t-t') \cdot
               \vec{E}^{ext}(\vec{r}',t'). \label{eq-1}
\end{eqnarray}
For simplicity we introduce a symbolic notation for the above equation in
the form $\vec j = \vec{\vec{\sigma}} \star \vec{E}^{ext}$.
In our work, the role of the external field  is taken by the auxiliary field 
$\vec{E}^{aux}$ that is homogeneous and has only a finite duration, i.e. 
it is absent for large times and it bears no information about the drop 
in potential in the long-time limit. The latter is contained within 
the {\it induced} field 
$\vec{E}^i(\vec{r},t)$, which appears explicitly when considering 
the irreducible (or proper) conductivity~\cite{Pines67,GuilianiVignaleBook},
\begin{equation}
\vec j = \vec{\vec{\sigma}} \star \vec{E}^{aux}=
\vec{\vec{\sigma}}^{irr} \star ( \vec{E}^{aux} + \vec{E}^i ). \label{eq-2}
\end{equation}
The induced field accounts for the electron-electron interaction
at the Hartree level, which is usually referred to as the long-range effects 
of the Coulomb interaction, whereas the rest of the interactions between 
electrons, the short range part, is included within the irreducible 
conductivity.
The quantity in which we are primarily interested in is the conductance, 
defined as $G = I/V$ where $V$ is a voltage drop. For $G$, the detailed 
spatial and time dependence of the current density and the induced
field while the steady current is being established are of no relevance.
We define the voltage drop as the overall drop in potential of the induced
field $\vec{E}^i$ along the current flow (along the $z$ axis) between 
far left and far right,
\begin{eqnarray}
             V = \lim_{t\rightarrow \infty} \int_{-\infty}^{+\infty}
             dz E_z^i(\vec{r},t), \label{eq-3}
\end{eqnarray}
which will be indicated by a superscript $4P$ in the associated
4-point conductance, $G^{4P}$. This definition is most suitable
\begin{figure}
\includegraphics[width=8cm]{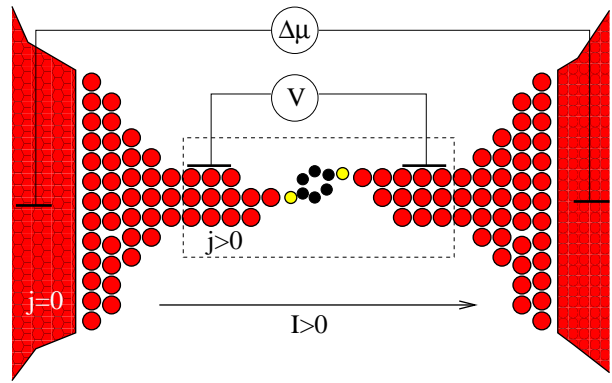}
\caption{(Color-online) Measurement of the voltage between 
the macroscopic electrodes, where the current density is zero, gives 
the two-point conductance $G^{2P}=I/(\Delta \mu/e)$. Measuring the 
voltage drop inside the simulation box (supercell), where the 
current density is nonzero, gives the four-point conductance $G^{4P}=I/V$.
Increasing the supercell the two quantities approach each other 
(see Sec.~\ref{sec-5}).}
\label{fig-0}
\end{figure}
for {\it ab-initio} modeling and does not suffer from ambiguities present
when one considers two chemical potentials. In mesoscopic physics, the above
definition of conductance is referred to as the 4-point conductance (hence
the superscript) since it represents measurement of the voltage drop
using contacts different from those acting as a source and drain of
the current. In the extreme case of a 1D conducting channel of non-interacting,
but locally neutral electrons it reduces to the well-known expression
$G^{4P} = (2e^2/h)(T/R)$.  This is in contrast with the most
frequently encountered 2-point conductance, $G^{2P}$, where the voltage drop
is understood to be the difference in electro-chemical potentials of
the two macroscopic electrodes, $\Delta \mu/e$ (see Fig.~\ref{fig-0}). 
Since the electrodes
are not part of the quantum-mechanical model of the conducting channel,
the familiar Landauer result $G^{2P}=(2e^2/h)T$ is believed not to
be derivable from the Kubo formalism~\cite{MahanBook} and its plausible derivations
are always accompanied with steps motivated by physical insight
and arguments about phase-incoherent adiabatically widening
electrodes~\cite{Landauer90a}.
Here we show that for a nanocontact between
massive electrodes the 4-point conductance approaches the 2-point Landauer
formula. However we never make a reference to $\Delta \mu/e$ and we
consistently work with the drop in the induced electrostatic potential only.
Our claims for its applicability for the experimental 2-point conductance
follow from the fact that for massive electrodes connected
with a nanojunction the current density in the electrodes goes to zero
and thereby the drop in the induced (Hartree) potential is an excellent
indicator
of the phenomenological quantity $\Delta \mu/e$.  The latter argument
serves as a motivation to use the $G^{4P}$ as conceptually and  
formally well defined quantity to characterize systems of fully
interacting electrons in any geometry of the junction.

\section{Singularity in the response function and the 2-point conductance}
\label{sec-3}
An essential property of the conductivity response function of extended
systems capable of coherent transport is their singular long range character
for large times or -- equivalently -- as the frequency approaches
zero~\cite{Bokes04}. To show this behavior explicitly we need to introduce
a few geometrical details of the studied nanojunction which do not restrict
the generality of our argument. We consider a geometry where $z$ is
a direction of the current flow, $q$ a reciprocal vector in that direction,
and for clarity we assume that the system can be put into a periodic
supercell along the $x,y$ directions. The conductance $G^{4P}$ is then
a conductance for the area of the supercell $S$.
To find the current flowing through $S$, we integrate
Eq.~(\ref{eq-2}) across the cross-sectional area. This naturally 
leads to the cross-section integrated conductivity~\cite{Sols91}
\begin{equation}
    \sigma^{irr}(z,z';t) = \int_{S}\int_{S} d\vec{S} \cdot  \label{eq-4}
            \vec{\vec{\sigma}}^{irr}(\vec{r},\vec{r}';t)
            \cdot d\vec{S}',
\end{equation}
which relates the current to the $z$-component of the total field,
\begin{equation}
    I(z,t) = \int^t dt' \int_{-\infty}^{\infty} dz'  \label{eq-5}
        \sigma^{irr}(z,z';t-t') ( E^{aux}(t') + E^{i}(z',t') ).
\end{equation}
The singular long-range character appears as the independence 
of $\sigma^{irr}(z,z';t)$ on $z$ and $z'$ as $t \rightarrow \infty$ 
which when performing the Fourier transforms
$z,z' \rightarrow q,q'$ and $t \rightarrow \omega+i\alpha$, takes the 
form~\cite{Bokes04}
\begin{equation}
    \lim_{\omega \rightarrow 0} \sigma^{irr}(q,q',\omega+i\alpha) =
    2\pi G^{2P} \delta(q) \delta(q'), \label{eq-6}
\end{equation}
where we have formally introduced the 2-point conductance, $G^{2P}$,
as the strength of the above mentioned singular character. To extract
the strength we introduce a linear functional $\mathcal{G}^{\sigma}[]$
\begin{equation}
    \mathcal{G}^{\sigma}[\sigma^{irr}] =
        \lim_{\alpha \rightarrow 0^+}  \label{eq-6.1}
        \int \int \frac{dq dq'}{2\pi} \sigma^{irr}(q,q';i\alpha),
\end{equation}
for which we simply have $G^{2P}=\mathcal{G}^{\sigma}[\sigma^{irr}]$.
It has been shown previously that this identification is correct
for noninteracting electrons~\cite{Bokes04}. Further motivation to refer to
it as a 2-point conductance for interacting systems will be discussed
in Section \ref{sec-5}.

\section{Onset of the current inside the electrodes}
\label{sec-4}
The Eq.~(\ref{eq-5}) can now be analyzed in the light of the above observations.
First, let us consider a distant part of one of the electrodes, 
$|z| \gg 0 $, for times such that $1/E_F \ll t \ll |z|/v_F$, where
$E_F$ and $v_{F}$ are the Fermi energy and Fermi speed respectively. In this
regime, the response in the electrodes 
($|z| \gg 0$) is determined by the electrode
itself, independently of the nanojunction. After a short relaxation time ($\sim 1/E_F$)
the auxiliary field establishes a uniform (i.e. $z$-independent) current inside the electrode
\begin{equation}
    I(z,t) = \int dz' \int_{-\infty}^t dt' \sigma^{irr,e}(z,z';t-t')
        E^{aux}(t'), \label{eq-7}
\end{equation}
where $\sigma^{irr,e}$ is the irreducible conductivity of the electrode 
only\footnote{For simplicity of presentation we assume that 
the induced field inside the electrode is absent, as is exactly true 
for jellium electrodes, used later in Sec.~\ref{sec-7}. Generalization 
for an electrode with atomic structure is straightforward: The induced 
field $\vec{E}^i$ is split into the local field induced inside the electrode, 
$\vec{E}^{i,e}$ (which has an average value zero over the electrode's unit 
cell), and the difference $\vec{E}^{i,j} = \vec{E}^i-\vec{E}^{i,e}$. 
The latter then defines the potential drop according to Eq.~\ref{eq-3} 
and the former needs to be included into the response entering 
the functional $\mathcal{F}^{\sigma}[\tilde{\sigma}]$ where 
$\tilde{\sigma}=\sigma\star(1 + \chi^e)$ with 
$\chi^e=\delta E^{i,e}/\delta E^{aux}$.}
while preserving the local charge neutrality within the electrode.
We are free to choose any time-dependence of the field as long as it
delivers finite change of momentum for each electron and thus establishes the eventual steady current; the most convenient
choice is $E^{aux}(t) = - A^{aux} \delta(t)$, where $A^{aux}$ is the magnitude
of the change in the homogeneous vector potential at $t=0$.

This picture is not valid as one approaches the region of the nanojunction,
i.e. for $z\sim0$. Still, since the current will keep coming from the left electrode
and disappearing into the right electrode (there is always such $|z|\gg0$ that
for any time $t$ we have $t\ll |z|/v_F$ and hence the above analysis can be applied),
the charge will have to pile up in front of the junction and similarly deplete behind
the junction, creating a resistivity dipole~\cite{Landauer57}. The incoming
current cannot be decreased by some form of reflected front/disturbance
from the junction since this would break the local charge neutrality
within the electrode, leading to strong opposing fields. 
Hence the dipole around the junction and therefore
the drop in the induced potential $V$ will grow, increasing the current in the junction region until
it becomes equal to the current, $I$,
deep inside the electrode, as described in the preceding paragraph.
This will be possible since in the linear
regime we expect $I=G^{4P} V$.

Having in mind this physical process, the established uniform current
at long times expressed by Eq.~(\ref{eq-7}) for the distant electrode applies at {\it all} positions including $z=0$.  
We express the electrode conductivity $\sigma^{irr,e}$ in reciprocal space for long times as
\begin{eqnarray}
    I(t\rightarrow \infty) &=& - \lim_{\alpha \rightarrow 0^+}
    \int dq \sigma^{irr,e}(q,q'=0;i\alpha) A^{aux}  \label{eq-8} \\
    &=& \mathcal{F}^{\sigma}[\sigma^{irr,e}] A^{aux}, \label{eq-9}
\end{eqnarray}
where we use a small imaginary frequency $i\alpha$ to perform the long-time
limit~\cite{Kubo59}. The linear functional $\mathcal{F}^{\sigma}[]$
will be further discussed in Section \ref{sec-7}.

\section{The 4-point formulation of the conductance}
\label{sec-5}
The above physical picture can be directly used within Eq.~(\ref{eq-5}).
We write the conductivity of the whole system as the conductivity
of an infinitely long electrode alone, $\sigma^{irr,e}$, plus a further term, $\sigma^{irr,j}$, characterizing
the presence of the junction: $\sigma^{irr}=\sigma^{irr,e}+\sigma^{irr,j}$. Eq.~(\ref{eq-5})
is thereby cast into the form
\begin{eqnarray}
    I &=& \sigma^{irr,e} \star E^{aux} + \sigma^{irr,j} \star E^{aux}
        + \sigma^{irr} \star E^i. \label{eq-9.1}
\end{eqnarray}
Evaluating this equation at $z=0$ and using the definitions of the functionals $\mathcal{F}^{\sigma}$
and $\mathcal{G}^{\sigma}$, the equation (\ref{eq-9.1}) can be written as
\begin{eqnarray}
I &=&  \mathcal{F}^{\sigma}[\sigma^{irr,e}] A^{aux} + \mathcal{F}^{\sigma}[\sigma^{irr,j}] A^{aux} + \mathcal{G}^\sigma[\sigma^{irr}] V.
\end{eqnarray}
From Eq.~(\ref{eq-7}) we know that in the long-time limit, the first  
term on the RHS is itself equal to $I$, so that the  second and third terms sum to zero,
from which we easily arrive at the central result of this paper
\begin{equation}
    G^{4P} = 
        \frac{\mathcal{F}^{\sigma}[\sigma^{irr,e}]}{
        \mathcal{F}^{\sigma}[\sigma^{irr,e}] -
        \mathcal{F}^{\sigma}[\sigma^{irr}]} \times
        \mathcal{G}^\sigma[\sigma^{irr}], \label{eq-10}
\end{equation}
where we have used the fact that
$\mathcal{F}^{\sigma}[\sigma^{irr,j}] =
\mathcal{F}^{\sigma}[\sigma^{irr}] - \mathcal{F}^{\sigma}[\sigma^{irr,e}]$
and that $V = \int dz E^i(z) = E^i(q=0)$.

It is very instructive to apply the above result for a simple 1D non-interacting gas
with a single scattering center\footnote{From here on we use the atomic units with $\hbar=m_e=e=1$.}.
For this system, the noninteracting
conductivity in the small frequency limit has a form~\cite{Kamenev01}
\begin{eqnarray}
    \sigma^0(q,q';i\alpha) &=& \frac{2}{\pi} \frac{v_F \alpha}{v_F^2 q^2 + \alpha^2}
        \delta(q-q') \nonumber \\
        && - R \frac{2}{\pi} \frac{v_F \alpha}{v_F^2 q^2 + \alpha^2}
               \frac{2}{\pi} \frac{v_F \alpha}{v_F^2 (q')^2 + \alpha^2}
\end{eqnarray}
The irreducible conductivity of the homogeneous 1D gas, corresponding here
to the electrode in the general case is simply the first part of the above
expression
\begin{equation}
    \sigma^{0,e}(q,q';i\alpha)  =
        \frac{2}{\pi} \frac{v_F \alpha}{v_F^2 q^2 + \alpha^2} \delta(q-q').
\end{equation}
It is straightforward to see that using the above forms we have
$\mathcal{G}^\sigma[\sigma^0]=(1-R)/\pi=T/\pi$,
$\mathcal{F}^\sigma[\sigma^{0,e}] \sim v_F/\pi \alpha$ and
$\mathcal{F}^\sigma[\sigma^{0,j}] \sim R v_F/\pi \alpha$,
and we arrive at the 4-point conductance $G^{4P}= (1/\pi)(T/R)$.
The arguments of local charge neutrality in the electrodes and its
consequences are very closely related to the original comment by
Thouless~\cite{Thouless81} and a more recent work by Kamenev and
Kohn~\cite{Kamenev01}. In the form presented in Eq.~(\ref{eq-10})
it is valid for nanojunctions of any shape as long as we can put them into
a supercell, including the extreme case of a planar metal-vacuum-metal
junction for which we will demonstrate its applicability in
Section~\ref{sec-7}.

The generality of Eq.~(\ref{eq-10}) can be used to show the equivalence
of the 2-point and 4-point conductances for nanojunctions with massive
electrodes. The demonstration is based on the observation that
for such a geometry $\mathcal{F}^{\sigma}[\sigma^{irr,e}] \gg
\mathcal{F}^{\sigma}[\sigma^{irr}]$ and hence the prefactor containing
the $\mathcal{F}[]$s in Eq.~(\ref{eq-10}) goes to 1. The mentioned
inequality is immediately clear from the following argument:
$\mathcal{F}^{\sigma}[\sigma^{irr,e}]$ gives the current
flowing in the electrodes when the vector potential is changed from
zero to $A^{aux}$, whereas $\mathcal{F}^{\sigma}[\sigma^{irr}]$
formally
gives current as a response to the same disturbance but in a fictitious
system having the same geometry as the real nanojunction but for which
the long-range Coulomb interaction is missing. This implies that
the local charge neutrality is not enforced (since only the irreducible
response enters $\mathcal{F}^{\sigma}$) and as a result most of the electrons
coming from one electrode will be reflected from the nanojunction and decrease
the current. Hence the resulting current will be much smaller
in this second case and the inequality is fulfilled. This formally
justifies Landauer's arguments in favor of the expression $G=T/\pi$ as
the conductance of a nanojunction between two phase-randomizing
and adiabatically widening reservoirs. As we can clearly see, it is only the
widening that is really needed to have the conductance of the
junction as a whole be given by Eq.~(\ref{eq-6.1}). This explains our
motivation to refer to the latter quantity as a 2-point conductance
even for interacting electronic systems. The argument applies also to
opaque barriers, not necessarily having a constriction, which has been
explored in previous works~\cite{Mera05,Carva04} and is also demonstrated
in Section \ref{sec-7}.

\section{Inclusion of exchange and correlation into conductance}
\label{sec-6}
Having established the validity and generality of Eq.~(\ref{eq-10})
we can proceed to approximations that go beyond non-interacting or
Hartree-like interacting electrons. The simplest way to do so
is within the framework of time-dependent density-functional
theory. Some doubt about the applicability of the latter might
arise from the dynamics deep in the electrode described in Section~\ref{sec-4}.
There, the induced current is dominated by a divergenceless component
that cannot be related to a time-dependent density, which is the only
physically relevant quantity within the TDDFT. However, at the level
of Fermi-liquid theory, this divergenceless current is identical
to that of a non-interacting system due to the backflow
of quasiparticles~\cite{NozieresBook}. Hence for this part we do not
expect corrections arising from the exchange and correlation and
we should expect $\mathcal{F}^\sigma[\sigma^{irr,e}]=
\mathcal{F}^\sigma[\sigma^{0,e}]$, where $\sigma^{0,e}$ is the non-interacting 
conductivity of the electrode. On the other hand, the response described
by the rest of Eq.~(\ref{eq-9.1}), leading to the voltage drop
in the induced potential, is essentially localized around the nanojunction
and is completely described by the time- and space- dependence
of the electronic density which is amenable to the TDDFT approach.

To cast our theory into the TDDFT framework, we need to reformulate
the functionals $\mathcal{F}^\sigma$ and $\mathcal{G}^\sigma$ since
the conductivity is not directly accessible within TDDFT.
The irreducible conductivity is simply related to the irreducible
polarization~\cite{Bokes04,Pines67}
\begin{eqnarray}  \label{eq-sigma-chi}
    \sigma^{irr}(q,q',i\alpha) = - \frac{\alpha}{qq'} \chi^{irr}(q,q';i\alpha),
\end{eqnarray}
which is conveniently calculated via the density response function calculation
within TDDFT. To incorporate this we also introduce new functionals of the
irreducible polarization
\begin{eqnarray}
    \mathcal{F}[\chi^{irr}] &=&
    \mathcal{F}^\sigma[ - \frac{\alpha}{qq'} \chi^{irr}(q,q';i\alpha) ], \\
    \mathcal{G}[\chi^{irr}] &=&
    \mathcal{G}^\sigma[ - \frac{\alpha}{qq'} \chi^{irr}(q,q';i\alpha) ].
\end{eqnarray}
for which we give explicit expressions suitable for direct numerical
evaluation in reciprocal and real space representation in the Appendix.
The irreducible polarization satisfies the Dyson equation
\begin{eqnarray}
    \chi^{irr}(q,q';i\alpha) = \chi^0(q,q';i\alpha) + \nonumber \\
        \int dq'' dq'''
        \chi^{0}(q,q'';i\alpha) f_{xc}(q'',q''';i\alpha)
        \chi^{irr}(q''',q';i\alpha) \label{eq-chi-dyson}, \label{eq-12}
\end{eqnarray}
where $\chi^0(q,q';i\alpha)$ is the non-interacting Kohn-Sham density response function,
and $f_{xc}(q'',q''';i\alpha)$ is a non-local frequency-dependent exchange-correlation
kernel~\cite{Vignale96}. The above two equations give the 4-point conductance of
an interacting electronic system: for a given kernel $f_{xc}$ one needs to invert
the Dyson equation (\ref{eq-12}), substitute the result into Eq.~(\ref{eq-sigma-chi})
and employ the general expression Eq.~(\ref{eq-10}).

However, we can gain more insight by multiplying Eq.~(\ref{eq-chi-dyson})
by $-\alpha/(qq')$ and taking the limit $\alpha \rightarrow 0$. Clearly,
the resulting left-hand-side is singular in $q$, with the strength being  
$G^{2P}$ according to Eq.~(\ref{eq-6}). The strength of the first
term on the right-hand-side of (\ref{eq-12}) multiplied by the same factor
gives the 2-point conductance of the non-interacting Kohn-Sham system,
$G^{2P,0}$.
The difference between these the two, i.e. the correction due
to the exchange-correlation kernel, is then nonzero \emph{only if
the last term in Eq.~(\ref{eq-chi-dyson}) also leads to a singular form}.
This observation can be used to deduce the forms of the kernel that \emph{do}
influence the conductance, since the general character of $\chi$ for
small $\alpha$ is well known.
The most interesting choice, making use of the character of
$\chi^{0/ir} \sim qq'$, is $f_{xc}(q,q';i\alpha) = \frac{\alpha}{qq'}
A(q,q'; i\alpha)$ where $A$ is well-behaved: $A(q,q';i\alpha) \rightarrow R^{\tt dyn}
= A(q=0,q'=0) \neq 0$ for $\alpha \rightarrow 0$. The resulting
4-point conductance then takes the form
\begin{eqnarray}
        G^{4P} &=& 
        \frac{\mathcal{F}[\chi^{0,h}]}{(1+\mathcal{G}[\chi^{0}] R^{\tt dyn})
    \mathcal{F}[\chi^{0,h}] - \mathcal{F}[\chi^{0}]}
    \mathcal{G}[\chi^{0}]. \label{eq-13}
\end{eqnarray}
In the case of a narrow junction between massive electrodes and
using the arguments leading to Eq.~(\ref{eq-10}), we obtain
\begin{equation}
        G^{4P} \approx 
        \frac{\mathcal{G}[\chi^{0}]}{1+\mathcal{G}[\chi^{0}] R^{\tt dyn} }
    = \frac{G^0}{1 + R^{\tt dyn} G^0}.
\end{equation}
i.e. in this very important case $R^{\tt dyn}$
represents part of the dynamical resistance, having origin
in exchange-correlation effects and can be accounted for by adding
this resistance in series with the Kohn-Sham result.

An approximate correction of this form has, in fact, been identified
by Na Sai {\it et al.}~\cite{Sai05}
for interacting electronic systems with weakly inhomogeneous potential along the
direction of the current flow.
For such systems it is possible to show that a purely longitudinal
exchange-correlation electric field used in their
treatment within TDCDFT~\cite{Vignale96,Vignale97}
is equivalent to a contribution to the TDDFT kernel of the asymptotic
form for small $q,q'$
\begin{eqnarray}
    f^{(\eta)}_{xc}(q,q';\omega) \approx
        - \frac{i\omega}{q q'} A \int dz \frac{4\eta}{3}
        \left( \frac{\partial_z n(z)}{n(z)^2}  \right)^2
    =    - \frac{i\omega}{q q'} A^2 R^{\tt dyn} \label{eq-A}
\end{eqnarray}
where $A$ is the cross-sectional area of the considered system, $n(z)$
is the electronic density and $\eta$ is the dynamical viscosity
of a homogeneous electron gas~\cite{Vignale97}. We should
note that for homogeneous systems $R^{\tt dyn}=0$ since $\partial_z n(z)=0$.
This is important since the functional form $f_{xc}^{(\eta)}$ given above
and the limiting process would not lead to a finite result for
a homogeneous system. It is interesting to note that the nonlocal
character of the kernel, i.e. $f^{xc} \sim 1/q^2$  has been found
also responsible for significant improvement in the {\it ab-initio} studies
of the optical spectra in many materials~\cite{Botti05,Reining02}.

A very popular approximation for the exchange-correlation kernel is
the adiabatic local-density approximation (ALDA)~\cite{Gross85}
\begin{eqnarray}
    f_{xc}^{(ALDA)}(\mathbf{r},\mathbf{r}') =
        f_{xc}^{ALDA}[n^0(\mathbf{r})]
        \delta (\mathbf{r}-\mathbf{r}') \\
    f_{xc}^{ALDA}[n^0]= d^2 ( n^0 \epsilon_{xc}(n^0) ) / dn^2,
    \label{eq-8o}
\end{eqnarray}
(Here $\epsilon_{xc}(n^0)$ is the exchange-correlation energy per particle of a
homogeneous electron gas (HEG) of density $n^0$), leading
to a kernel in the reciprocal space of the form
$f_{xc}^{(ALDA)}(q,q';\omega) = f_{xc}^{(ALDA)}(q-q';\omega) \rightarrow
B(\omega) \delta(q-q'), B(\omega) \rightarrow B \neq 0$ for
$\omega \rightarrow 0$. It has been argued by several authors that ALDA
should not contribute to any change in the
conductance.~\cite{Sai05,Burke05}
The basis of this argument comes from the fact that one can
account for the exchange and correlation effects either via
the exchange correlation kernel, or by using the exchange-correlation
electric field. The latter can be shown to contribute to the conductance
only if there is a nonzero drop of the
exchange-correlation
potential across the system, which is absent within any local or semi-local approximations.
The argument is valid also within our theory when applied to
Eq.~(\ref{eq-9.1}). However, it is interesting to explore
the consequences if one stays within the kernel-based treatment. Here
the proper definition of the 4-point conductance is essential. If one
directly uses the 2-point formula in Eq.~(\ref{eq-6.1}) for metal-vacuum-metal
interfaces, one finds nontrivial ALDA corrections solely due to the fact
that the Kohn-Sham conductivity does redistribute the charge
at long distances when a localized Kohn-Sham potential with nonzero drop
is applied~\cite{Bokes_condmat06}. The numerical results
for metal-vacuum-metal presented in Section \ref{sec-7} show
a cancellation between these ALDA corrections within
$\mathcal{G}[\chi^{irr}]$ and those entering $\mathcal{F}[\chi^{irr}]$
within the precision of the numerics so that the $G^{4P}$ coming from Eq.~(\ref{eq-10}) remains unaffected
by the ALDA kernel, 
as it should be.
It is important to emphasize this point
because studies of exchange and correlation effects {\it using the 2-point
formula only} for 1D atomic chains~\cite{Prodan07} or
quantum wires~\cite{Malet05} may easily lead to incorrect conclusions.

\section{Conductance of a metal-vacuum-metal interface}
\label{sec-7}
The metal-vacuum-metal junction is one of the simplest systems for {\it ab-initio}
study of electronic transport and hence it conveniently serves as a  
demonstration that the formula (\ref{eq-10}) can be numerically
implemented even for realistic systems. In practical implementations
we cannot work with an infinite system, for which the response function
that enters the functionals $\mathcal{F}[~]$ and $\mathcal{G}[~]$ is needed.
Instead, we take a finite system with $z\in(-L,L)$ with zero
boundary conditions at the end points. This brings certain restriction
on the zero-frequency extrapolation that we will discuss below.
The two perpendicular directions can be dealt with easily in the reciprocal
space for this particular system.

In our calculations we employ two jellium slabs of thickness $l$ and
density given by $r_s=3.0$~a.u., separated by a distance $d$,
i.e. $2l+d < 2L$ where the supercell extends from $-L$ to $L$. The calculation of $\chi^0(z,z';i\alpha)$ is performed
at the self-consistent LDA level~\cite{Jung04}. Subsequently for
the ALDA calculation we invert
the Dyson equation, Eq.~(\ref{eq-4}), in real space and thereby calculate
the irreducible response $\chi^{irr}(z,z';i\alpha)$. The inclusion of
the non-local kernel, such as that arising from the viscosity of the
electron liquid, can be done directly via Eq.~(\ref{eq-13}), avoiding
the inversion of the Dyson equation. For the evaluation of $\mathcal{F}[~]$
and $\mathcal{G}[~]$ we employ their real space form (Appendix),
integrating over the simulation cell,
\begin{eqnarray}
    \mathcal{G}_\alpha[\chi^{irr}] &=&
        \alpha \int_{-L}^0 \int_0^L dz dz'
        \chi^{irr}(z,z';i\alpha), \label{eq-G-2P-num}  \\
    \mathcal{F}_\alpha[\chi^{irr}] &=&
        \alpha \int_{-L}^{0} dz \int_{-L}^{L}  dz' \chi^{irr}(z,z') z',
\end{eqnarray}
keeping $\alpha$ finite for the moment. For extrapolation we need to
use imaginary frequencies $\alpha \ll E_F$ so that the transient dynamics is removed
from the response function. On the other hand, the finite extent
of the electrodes restricts the limiting procedure for
$\alpha \rightarrow 0$; one can go down no more than to imaginary
frequencies $\alpha_{min}\approx v_{F}/l$
($v_F\approx 0.64$ a.u. for $r_s=3$ a.u.).

\begin{figure}
\includegraphics[width=8cm]{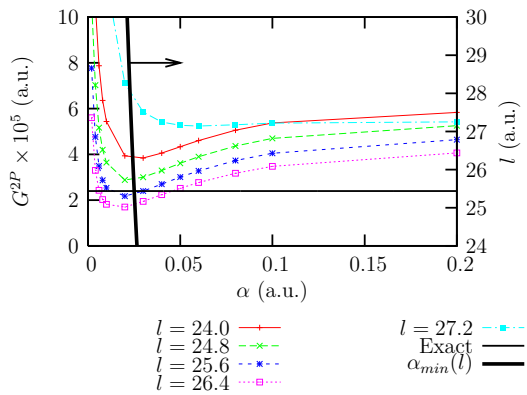}
\caption{(Color-online)
Dependence of $G^{2P}(\alpha)=\mathcal{G}_{\alpha}[\chi^0]$ on
frequency $\alpha$
for several electrode widths $l$ (vacuum is fixed for $d=8$~a.u.).
For frequencies below the thick line, given by $\alpha_{min}(l)=v_F/l$,
the finite size effects appear, so that $\alpha\sim0.05$ are used
for extrapolation to zero. The vertical shifts of the individual curves
is caused by the oscillation of the Fermi energy (see also Fig.~\ref{fig-2}).}
\label{fig-1}
\end{figure}

To demonstrate the precision and scaling of our calculation we first
present calculations for a non-selfconsistent square-barrier
potential (width $8$~a.u., height $0.25$ a.u.) between two
3D electrodes ($E_F=0.2$ a.u.). The
chosen values are reasonably close to our self-consistent potential
but at the same time allow for comparison of with the exact value of
$G^{2P} = \lim_{\alpha \rightarrow 0} \mathcal{G}_{\alpha}[\chi^0]$ using
the analytically known form of the transmission probability.
The calculated values of $\mathcal{G}_{\alpha}[\chi^0]$ shown in the
Fig.~\ref{fig-1} for finite slabs and finite frequencies clearly
show finite size effects below $\alpha_{min}$.
However, even for frequencies larger than $\alpha_{min}$ we observe
a remanent horizontal oscillatory dependence of the conductance curves on
the electrodes' width.
\begin{figure}
\includegraphics[width=8cm]{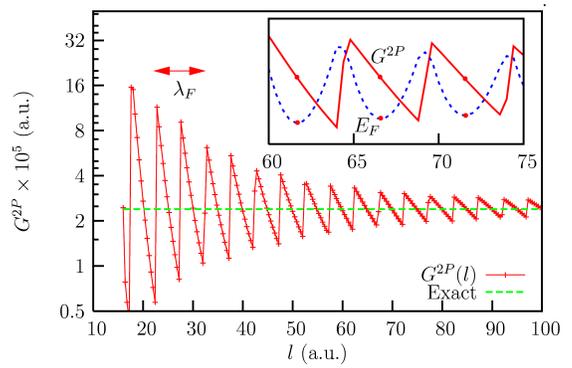}
\caption{(Color-online)
Dependence of the extrapolated $G^{2P}$ on the electrodes' widths $l$
for $d=8.0$ a.u for a square-barrier potential. The inset shows the oscillations in the Fermi
energy and the conductance $G^{2P}$ in a restricted range of $l$ for a self-consistent calculation. Choosing widths for which the Fermi energy attains its minima
(indicated with dots) leads to a stable estimate of the conductance of the infinite system.}
\label{fig-2}
\end{figure}
Even though the amplitude of the oscillations for smaller widths
are substantial, the overall convergence to the exact value is evident.
The oscillatory character of the Fermi energy can be effectively used
for identifying the most suitable system widths for an optimal estimate
of the infinite system conductance. One should choose the widths for which
the Fermi energy attains local minima (inset in Fig.~\ref{fig-2}),
which is the method we use within our work.

Similar extrapolation to zero frequency can be done for
$\mathcal{F}[~]$, or even better for $\alpha \times \mathcal{F}[~]$ which
approaches a constant value for an infinite system. However,
since our aim is to calculate the 4-point conductances, we directly
extrapolate the expression (\ref{eq-13}) which behaves very similarly
as the $G^{2P}(\alpha)$ described in detail in the preceding paragraph.

\begin{figure}
\includegraphics[width=8cm]{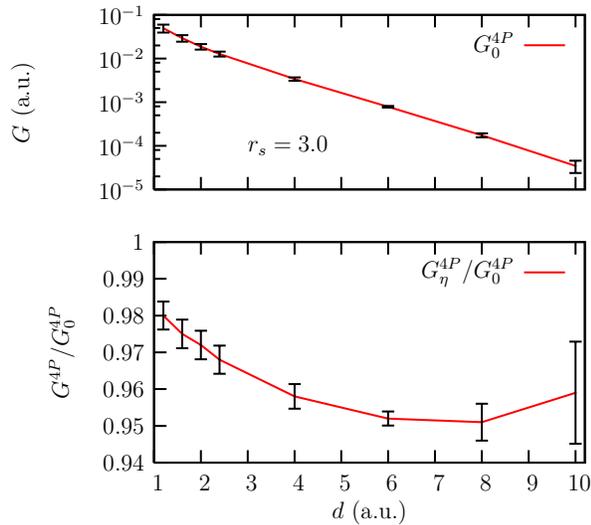}
\caption{(Color-online)
The Kohn-Sham conductance $G_0^{4P}$ with error bars resulting from uncertainty in the extrapolation to zero frequency (upper panel)
and the viscosity-corrected conductance $G_{\eta}^{4P}$ (shown relative to $G_0^{4P}$) (lower panel),
as a function of the vacuum width $d$. Only the corrections due to
the non-local kernel lead to changes in the conductance. The
ALDA kernel does not affect the Kohn-Sham result within numerical error.}
\label{fig-3}
\end{figure}

We now apply our methodology to a fully self-consistent calculation
for a sequence of vacuum widths $d=1,\ldots,10$ a.u. to study the effect
of the recently suggested viscosity-related exchange-correlation
correction due to Na Sai {\it et al.}~\cite{Sai05}, and to explore the cancellation
of ALDA corrections between $\mathcal{F}[~]$ and $\mathcal{G}[~]$.
The overall dependence of the conductance on $d$ is exponential.
For clarity of presentation we show first the exponentially
decreasing form of the calculated Kohn-Sham conductance, $G_0^{4P}$, in the upper panel
of Fig.~\ref{fig-3};
the small symbols indicate error bars in $G_0^{4P}$ due to uncertainty
in the extrapolation to zero frequency.
For this system the absolute values of the corrected conductances after inclusion of either TDDFT kernel
-- ALDA kernel or viscosity kernel -- lie within these indicated error bars.
As we have discussed in Section~\ref{sec-6}, the ALDA does not lead to any correction, but
numerically this result is not trivial. In fact both, the $\mathcal{G}[\chi^{ALDA}]$ and
$\mathcal{F}[\chi^{ALDA}]$ do change due to the presence of the
exchange-correlation kernel (\ref{eq-8o}), but these changes are
canceled in the total expression for the 4-point conductance
(\ref{eq-10}) within the numerical precision of the extrapolation to zero frequency.
On the other hand, the corrections due to the finite viscosity of the electron liquid {\it do} lead
to a small but systematic decrease in the conductance, which is about 5\% in the examined
range of relative conductances~\cite{Sai05,Jung07}. While this is smaller than the
error bar of the extrapolated conductance, since the viscosity correction itself does not involve
extrapolation (Eq.~\ref{eq-8o}), the resulting error bar of the {\it relative} correction,
shown in the lower panel of Fig.~\ref{fig-3}, is noticeably smaller than the correction itself, and thus the
correction is significant.

\begin{figure}
\includegraphics[width=8cm]{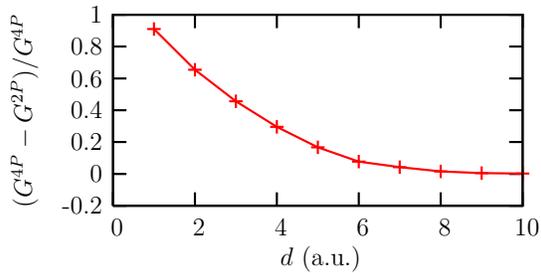}
\caption{(Color-online)
The relative difference between the 4-point and 2-point Kohn-Sham conductances.
The difference between the two diminishes for opaque barriers.}
\label{fig-4}
\end{figure}

Finally we numerically demonstrate the appropriateness of the
2-point conductance, $G^{2P}=\mathcal{G}[\chi]$ for opaque barriers.
The graph in Fig.~\ref{fig-4} clearly shows that for $d>4$ the
$\mathcal{F}[~]$-dependent prefactor is very close to 1, which supports
our arguments in Sec.~\ref{sec-5} as well as is in agreement with previous
numerical work~\cite{Mera05}

\section{Conclusion}
In conclusion, we have presented a unified formalism, based
on singular character of response functions, that gives the
conductance of a general system of interacting electrons.
We have derived a closed formula for the 4-point conductance
in terms of the many-body response functions: the irreducible conductivity
or the irreducible density response. The formulation allows for
clear demonstration of the validity of the Landauer formula
for broadening or opaque nanojunctions. Furthermore, we have
utilized our formulation for examining the exchange and correlation
effects on the conductance within the time-dependent density functional
theory. We have shown that the long time limit determines
the functional form of the exchange-correlation kernel that can lead
to nonzero corrections. The exposed theory can be also used for {\it ab-initio}
calculations for which the achievable precision in conductance calculation
for a given size of a finite-size simulation cell has been demonstrated
on a simple metal-vacuum-metal junction.

\begin{acknowledgments}
This research was supported by
the Slovak grant agency VEGA (project No. 1/2020/05),
the NATO Security Through Science Programme (EAP.RIG.981521) and the EU's 6th
Framework Programme through the NANOQUANTA Network of Excellence
(NMP4-CT-2004-500198).
\end{acknowledgments}

\section{Appendix: Evaluation of functionals $\mathcal{F}$ and $\mathcal{G}$.}

The functionals $\mathcal{F}[~]$ and $\mathcal{G}[~]$ were given explicitly
in the Fourier-transformed form in Eqs.~(\ref{eq-8}),~(\ref{eq-9})
and Eq.~(\ref{eq-6.1}) respectively. In numerical calculations it is
more advantageous and numerically stable to use their representation
in real space. To achieve this, we use the inverse transform
of the irreducible density response function defined by
\begin{equation}
    \chi(q,q') =  \int \frac{dz dz'}{2\pi} e^{-iqz}
    \chi(z,z') e^{iq' z'}. \label{eq-app-1}
\end{equation}
Several times we will have to resolve the integral of the type
\begin{equation}
    f(z_0) = \int \frac{dq}{q} e^{iq z_0} \chi(q,q').
\end{equation}
Since $\chi(q,.) \sim q$ for small $q$, the integral is well defined
and we can choose to interpret the apparent singularity $1/q$ as
$1/(q+i\delta)$ or $1/(q-i\delta)$ with $\delta \rightarrow 0^+$.
Taking the former (the final result is independent of this choice)
and using the inverse transform (\ref{eq-app-1}) we find
\begin{equation}
    f(z_0) = - i \int dz dz' \theta(z_0 - z)
        \chi(z,z') e^{iq'z'}, \label{eq-app-3}
\end{equation}
where $\theta(z)$ is the unit step function.

Using the definition of $\mathcal{G}[~]$ and using the integral
(\ref{eq-app-3}) twice with $z_0=0$ we readily obtain
\begin{equation}
    \mathcal{G}[\chi] =  \lim_{\alpha \rightarrow 0^+}
    \alpha \int_{-\infty}^0 dz \int_{0}^{\infty} dz'
        \chi(z,z';i\alpha).
\end{equation}
The real-space form of the $\mathcal{F}$ is obtained using
again the Fourier transform:
\begin{eqnarray}
    \mathcal{F}[\chi]  
    &=& \left. \alpha \int \frac{dq}{qq'} \int \frac{dz dz'}{2\pi}
        e^{-iqz} \chi(z,z') e^{iq'z'} \right|_{q'=0} \\
    &=& \left. - i \alpha \int_0^{\infty} dz
        \int_{-\infty}^{\infty} dz' \chi(z,z')
        \frac{e^{iq'z'}}{q'} \right|_{q'=0} \label{eq-R-1} \\
    &=& \alpha \int_{-\infty}^{0} dz \int_{-\infty}^{\infty}  dz'
        \chi(z,z') z'  
\end{eqnarray}
since we can Taylor expand $e^{-iq'z'}$ with the linear term giving
the only nonzero contribution.


\end{document}